\documentclass{article} 
\usepackage{nips10submit_e,times}
\nipsfinalcopy

\usepackage[psamsfonts]{amssymb}
\usepackage[pdftex]{graphicx}
\usepackage{sidecap}

\newtheorem{theorem}{Theorem}[section]

\title{Deciphering subsampled data: adaptive compressive sampling as a principle of brain communication}

\author{
Guy Isely \\
Redwood Center for Theoretical Neuroscience\\
University of California, Berkeley\\
\texttt{guyi@berkeley.edu} \\
\And
Christopher J. Hillar \\
Mathematical Sciences Research Institute \\
\texttt{chillar@msri.org} \\
\AND
Friedrich T. Sommer\\
University of California, Berkeley\\
\texttt{fsommer@berkeley.edu} \\
}

\begin{document}

\maketitle

\begin{abstract}
A new algorithm is proposed for a) unsupervised learning of sparse representations from subsampled measurements and b) estimating the parameters required for linearly reconstructing signals from the sparse codes. We verify that the new algorithm performs efficient data compression on par with the recent method of compressive sampling. Further, we demonstrate that the algorithm performs robustly when stacked in several stages or when applied in undercomplete or overcomplete situations. The new algorithm can explain how neural populations in the brain that receive subsampled input through fiber bottlenecks are able to form coherent response properties.
\end{abstract}

\section{Introduction}
In the nervous system, sensory and motor information, as well as internal brain states, are represented by action potentials in populations of neurons. Most localized structures, such as sensory organs, subcortical nuclei and cortical regions, are functionally specialized and need to communicate through fiber projections to produce coherent brain function \cite{fellemanVanEssen1992}. Computational studies of the brain usually investigate particular functionally and spatially defined brain structures. Our scope here is different as we are not concerned with any particular brain region or function. Rather, we study the following fundamental communication problem: How can a localized neural population interpret a signal sent to its synaptic inputs without knowledge of how the signal was sampled or what it represents?  We consider the generic case that information is encoded in the activity of a local population (e.g. neurons of a sensory organ or a peripheral sensory area) and then communicated to the target region through an axonal fiber projection. Any solution of this communication problem is constrained by the following known properties of axonal fiber projections:

\textbf{Exact point-to-point connectivity genetically undefined}: During development, genetically informed chemical gradients coarsely guide the growth of fiber projections but are unlikely to specify the precise synaptic patterns to target neurons \cite{wymanthomas1983}. Thus, {\it learning mechanisms} and synaptic plasticity seem necessary to form the precise wiring patterns from projection fibers to target neurons. 

\textbf{Fiber projections constitute wiring bottlenecks}:  The number of axons connecting a pair of regions is often significantly smaller than the number of neurons encoding the representation within each region \cite{schuzetal2006}. Thus, communication across fiber projections seems to rely on a form of {\it compression}. 

\textbf{Sizes of origin and target regions may differ}:  In general, the sizes of the region sending the fibers and the region targeted by them will be different. Thus, communication across fiber projections will often involve a form of {\it recoding}.

We present a new algorithm for establishing and maintaining communication that satisfies all three constraints above.  To model imprecise wiring, we assume that connections between regions are configured randomly and that the wiring scheme is unknown to the target region.  To account for the bottleneck, we assume these connections contain only subsampled portions of the information emanating from the sender region; i.e., learning in the target region is based on subsampled data and not the original. 

Our work suggests that axon fiber projections can establish interfaces with other regions according to the following simple strategy: {\it Connect to distant regions randomly, roughly guided by chemical gradients, then use local unsupervised learning at the target location to form meaningful representations of the input data.}  Our results can explain experiments in which retinal projections were redirected neonatally to the auditory thalamus and the rerouting produced visually responsive cells in auditory thalamus and cortex, with properties that are typical of cells in visual cortex \cite{sur1988}. Further, our model makes predictions about the sparsity of neural representations. Specifically, we predict that neuronal firing is sparser in locally projecting neurons (upper cortical layers) and less sparse in neurons with nonlocal axonal fiber projections. In addition to the neurobiological impact, we also address potential technical applications of the new algorithm and relations to other methods in the literature.

\section{Background}

\textbf{Sparse signals}:  It has been shown that many natural signals falling onto sensor organs have a higher-order structure that can be well-captured by sparse representations in an adequate basis; see \cite{ruderman1994, olshausenfield1996} for visual input and \cite{bellsejnowski1996, smithlewicki2006} for auditory.   The following definitions are pivotal to this work.  

{\it Definition~1:}  An ensemble of signals $X$ 
within  $\mathbb R^n$ has {\it sparse underlying structure} if there is a dictionary $\Omega \in \mathbb R^{n \times p}$ so that any point $\mathbf{x} \in \mathbb R^n$ drawn from $X$ can be expressed as $\mathbf x = \Omega \mathbf v$ for a  sparse vector $\mathbf v \in \mathbb R^p$. 


{\it Definition~2:} An ensemble of sparse vectors $V$ 
within $\mathbb R^p$ is a {\it sparse representation} of a signal ensemble $X$ in $\mathbb R^n$ if there exists a dictionary $\Omega \in \mathbb R^{n \times p}$ such that the random variable $X$ satisfies $X = \Omega V$. 


For theoretical reasons, we consider \textit{ensembles} of random vectors (i.e. random variables) which arise from an underlying probability distribution on some measure space, although for real data sets (e.g. natural image patches) we cannot guarantee this to be the case.  Nonetheless, the theoretical consequences of this assumption (e.g. Theorem \ref{mainPsiDiagPermThm}) appear to match what happens in practice for real data (figures 2-4).


\textbf{Compressive sampling with a fixed basis}: 
Compressive sampling (CS) \cite{cands2006} is a recent method for representing data with sparse structure using fewer samples than required by the Nyquist-Shannon theorem.  In one formulation \cite{wainwright2009}, a signal $\mathbf{x} \in \mathbb{R}^n$ is assumed to be \textit{$k$-sparse} in an $n \times p$ dictionary matrix $\Psi$; that is, ${\mathbf x} = \Psi {\mathbf a}$ for some vector ${\mathbf a} \in \mathbb R^p$ with at most $k$ nonzero entries.   Next, $\mathbf x$ is subsampled using an $m \times n$ incoherent matrix $\Phi$ to give noisy measurements ${\mathbf y} = \Phi {\mathbf x} + \mathbf{w}$ with $m \ll n$ and independent noise $\mathbf{w} \sim \mathcal{N}(0,\sigma^2 I_{m \times m})$.  To recover the original signal, the following convex optimization problem (called Lasso in the literature)  is solved:
\begin{equation}\label{Lasso}
{\widehat{\mathbf b}}({\mathbf y}) :=   \arg \min_{\mathbf{a}} \left\{ \frac{1}{2n} ||{\mathbf y} - \Phi \Psi \mathbf{b}||_2^2 + \lambda |\mathbf{b}|_1 \right\},
\end{equation}
and then ${\widehat{\mathbf x}} := \Psi \widehat{\mathbf b}$ is set to be the approximate recovery of $\mathbf x$.  Remarkably,  as can be shown using \cite[Theorem 1]{wainwright2009}, the preceding algorithm determines a unique $\widehat{{\mathbf b}}$ and is guaranteed to be exact within the noise range:
\begin{equation}\label{uptothenoise}
||{\mathbf x}- \widehat{{\mathbf x}}||_2 = O(\sigma)
\end{equation}
with high probability (exponential in $m/k$) as long as the matrix $\Phi \Psi$ satisfies mild incoherence hypotheses, $\lambda = \Theta(\sigma \sqrt{ (\log p) /m})$, and the sparsity is on the order $k = O(m/\log p)$.  

Typically, the matrix $\Psi$ is $p \times p$ orthogonal, and the incoherence conditions reduce to deterministic constraints on $\Phi$ only.  Although in general it is very difficult to decide whether a given $\Phi$ satisfies these conditions, it is known that many random ensembles, such as i.i.d. $\Phi_{ij} \sim \mathcal{N}(0,1/m)$, satisfy them with high probability.  In particular, compression ratios on the order  $(k \log p)/p$ are achievable for $k$-sparse signals using a random $\Phi$ chosen this way.  



\textbf{Dictionary learning by sparse coding}:  For some natural signals there are well-known bases (e.g. Gabor wavelets, the DCT) in which those signals are sparse or nearly sparse. However, an arbitrary class of signals can be sparse in unknown bases, some of which give better encodings than others.  It is compelling to learn a sparse dictionary for a class of signals instead of specifying one in advance.  Sparse coding methods \cite{olshausenfield1996} learn dictionaries by minimizing the empirical mean of an energy function that combines the $\ell_2$ reconstruction error with a sparseness penalty on the encoding:
\begin{equation}
E({\mathbf  x}, {\mathbf  a}, \Psi) = ||{\mathbf  x} - { \Psi} {\mathbf  a}||_2^2 + \lambda S({\mathbf  a}).
\label{SCObjFunc}
\end{equation}
A common choice for the sparsity penalty $S(a)$ that works well in practice is the $\ell_1$ penalty $S({\mathbf a}) = | {\mathbf a}|_1$.  Fixing $\Psi$ and $\mathbf x$ and minimizing (\ref{SCObjFunc}) with respect to $\mathbf a$ produces a vector $\widehat{\mathbf{a}}(\mathbf{x})$ that approximates a sparse encoding for $\mathbf x$.\footnote{As a convention in this paper, $\mathbf a$ vs. $\mathbf b$ denotes a sparse representation inferred from full vs. compressed signals.}   For a fixed set of signals $\mathbf x$ and encodings $\mathbf a$, minimizing the mean value of (\ref{SCObjFunc}) with respect to $\Psi$ and renormalizing columns produces an improved sparse dictionary.  Alternating optimization steps of this form, one can learn a dictionary that is tuned to the statistics of the class of signals studied.  Sparse coding on natural stimuli has been shown to learn basis vectors that resemble the receptive fields of neurons in early sensory areas \cite{olshausenfield1996,rehnsommer2007,rozelletal2008}. Notice that once an (incoherent) sparsity-inducing dictionary $\Psi$ is learned, inferring sparse vectors $\widehat{\mathbf{a}}(\mathbf{x})$ from signals $\mathbf{x}$ is an instance of the Lasso convex optimization problem.  


\textbf{Blind Compressed Sensing}:  With access to an uncompressed class of sparse signals, dictionary learning can find a sparsity-inducing basis which can then be used for compressive sampling.  But what if the uncompressed signal is unavailable?  Recently, this question was investigated in \cite{gleichmaneldar2010} using the following problem statement.


{\it Blind compressed sensing (BCS):}  Given a measurement matrix $\Phi$ and measurements $\{ \mathbf y_1,\ldots,\mathbf y_N\}$ of signals $\{\mathbf x_1, \ldots, \mathbf x_N\}$ drawn from an ensemble $X$, find a dictionary $\Psi$ and $k$-sparse vectors $\{ \mathbf b_1,\ldots,\mathbf b_N\}$ such that $\mathbf{x}_i = \Psi \mathbf{b}_i$ for each $i = 1,\ldots,N$.  

It turns out that the BCS problem is ill-posed in the general case \cite{gleichmaneldar2010}. The difficulty is that though it is possible to learn a sparsity-inducing dictionary $\Theta$ for the measurements $Y$, there are many decompositions of this dictionary into $\Phi$ and a matrix $\Psi$ since $\Phi$ has a nullspace.  Thus, without  additional assumptions, one cannot uniquely recover a dictionary $\Psi$ that can reconstruct $\mathbf{x}$ as $\Psi \mathbf{b}$.  

\section{Adaptive Compressive Sampling}

It is tantalizing to hypothesize that a neural population in the brain could combine the principles of compressive sampling and dictionary learning to form sparse representations of inputs arriving through long-range fiber projections. Note that information processing in the brain should rely on faithful representations of the original signals but does not require a solution of the ill-posed BCS problem which involves the full reconstruction of the original signals.  Thus, the generic challenge a neural population embedded in the brain might have to solve can be captured by the following problem.

\begin{SCfigure}[][h]
\includegraphics[keepaspectratio, width=2in]{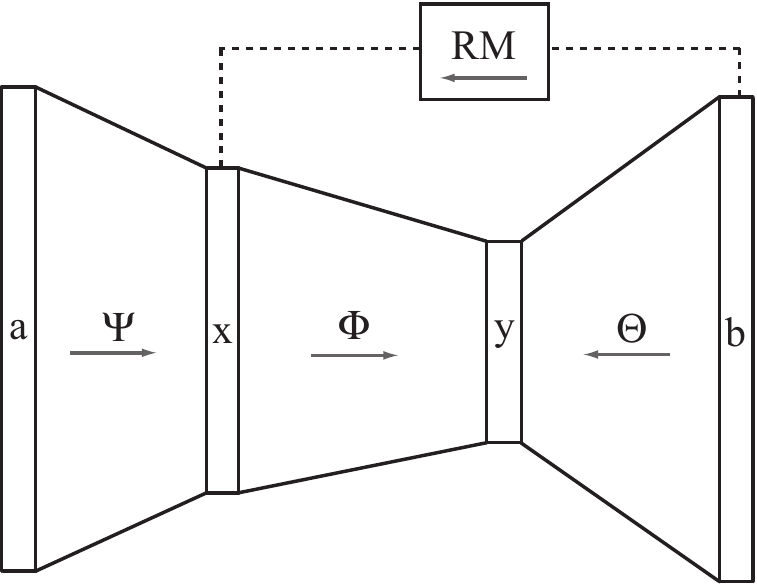}
\caption{ACS schematic.  A signal $\mathbf{x}$ with sparse structure in dictionary $\Psi$ is sampled by a compressing measurement matrix $\Phi$, constituting a transmission bottleneck. The ACS coding circuit learns a dictionary $\Theta$ for $\mathbf{y}$ in the compressed space, but can be seen to form sparse representations $\mathbf b$ of the original data $\mathbf x$ as witnessed by the matrix $RM$ in (\ref{RMeqn}). }
\label{fig1}
\end{SCfigure}

{\it Adaptive compressive sampling (ACS):} Given measurements $Y = \Phi X$ generated from an unknown $\Phi$ and unknown signal ensemble $X$ with sparse underlying structure, find signals $B(Y)$ which are sparse representations of $X$.

Note the two key differences between the ACS and the BCS problem. First, the ACS problem asks only for sparse representations $\mathbf b $ of the data, not full reconstruction.  Second, the compression matrix $\Phi$ is unknown in the ACS problem but is known in the BCS problem. Since it is unrealistic to assume that a brain region could have knowledge of how an efferent fiber bundle subsamples the brain region it originates from, the second difference is also crucial. We propose a relatively simple algorithm for potentially solving the ACS problem: use sparse coding for dictionary learning in the compressed space.  The proposed ACS objective function is defined as:
\begin{equation}\label{ACSobjective}
E({\mathbf y}, {\mathbf b}, \Theta) = ||{\mathbf y} - { \Theta} {\mathbf b}||_2^2 + \lambda S({\mathbf b}).
\end{equation}
Iterated minimization of the empirical mean of this function first with respect to $\mathbf{b}$ and then with respect to $\Theta$ will produce a sparsity dictionary $\Theta$ for the compressed space and sparse representations $\widehat{\mathbf b}(\mathbf y)$ of the $\mathbf y$.  Our results verify theoretically and experimentally that once the dictionary matrix $\Theta$ has converged, the objective (\ref{ACSobjective}) can be used to infer sparse representations of the original signals $\mathbf x$ from the compressed data $\mathbf y$. As has been shown in the BCS work, one cannot uniquely determine $\Psi$ with access only to the compressed signals $\mathbf y$. But this does not imply that no such matrix exists.  In fact, given a separate set of uncompressed signals $\mathbf x'$, we calculate a reconstruction matrix $RM$ demonstrating that the $\widehat{\mathbf b}$ are indeed sparse representations of the original $\mathbf x$.  Importantly, the $\mathbf x'$ are not used to solve the ACS problem, but rather to demonstrate that a solution was found.

The process for computing $RM$ using the $\mathbf x'$ is analogous to the process used by electrophysiologists to measure the receptive fields of neurons.  Electrophysiologists are interested in characterizing how neurons in a region respond to different stimuli.  They use a simple approach to determine these stimulus-response properties: probe the neurons with an ensemble of stimuli and compute stimulus-response correlations.  Typically it is assumed that a neural response $\mathbf b$ is a linear function of the stimulus $\mathbf x$; that is, $\mathbf b = RF \mathbf x$ for some receptive field matrix $RF$.  One may then calculate an $RF$ by minimizing the empirical mean of the prediction error:  $E(RF) = \| \mathbf{b} - RF \mathbf{x} \|_2^2$.  As shown in \cite{theunissen2001}, the closed-form solution to this minimization is $RF = C_{ss}^{-1} C_{sr}$, in which $C_{ss}$ is the stimulus autocorrelation matrix $\langle \mathbf{x} \mathbf{x}^{\top} \rangle_{X}$, and $C_{sr}$ is the stimulus-response cross-correlation matrix $\langle \mathbf{x} \mathbf{b}^{\top} \rangle_{X}$.  

In contrast to the assumption of a linear response typically made in electrophysiology, here we assume a linear generative model: $\mathbf x = \Psi \mathbf a$.  Thus, instead of minimizing the prediction error, we ask for the {\it reconstruction matrix} $RM$ that minimizes the empirical mean of the reconstruction error:
\begin{equation}
E(RM) = \| \mathbf{x} - RM \mathbf{b}  \|_2^2.
\end{equation}
In this case, the closed form solution of this minimization is given by
\begin{equation}\label{RMeqn}
RM = C_{sr} C_{rr}^{-1},
\end{equation}
in which $C_{sr}$ is the stimulus-response cross-correlation matrix as before and $C_{rr}$ is the response autocorrelation matrix $\langle \widehat{\mathbf{b}}(\mathbf{y}(\mathbf{x})) \widehat{\mathbf{b}}(\mathbf{y}(\mathbf{x}))^{\top} \rangle_{X}$.  As we show below, calculating (\ref{RMeqn}) from a set of uncompressed signals $\mathbf x'$ yields an $RM$ that reconstructs the original signal $\mathbf x$ from $\mathbf{\widehat{b}}$ as $\mathbf{x} = RM \mathbf{\widehat{b}}$.  Thus, we can conclude that encodings $\mathbf{\widehat{b}}$ computed by ACS are sparse representations of the original signals.  

\section{Theoretical Results}

The following hold for ACS under mild hypotheses (we postpone details for a future work).

\begin{theorem}\label{short_ver_expThm}
Suppose that an ensemble of signals is compressed with a random projection $\Phi$.   If ACS converges on a sparsity-inducing dictionary $\Theta$ and $C_{rr}$ is invertible, then $\Theta = \Phi \cdot RM$.
\end{theorem}

\begin{theorem}\label{mainPsiDiagPermThm}
Suppose that an ensemble of signals has a sparse representation with dictionary $\Psi$.   If ACS converges on a sparsity-inducing dictionary, then the outputs of ACS are a sparse representation for the original signals in the dictionary of the reconstruction matrix $RM$ given by (\ref{RMeqn}).  Moreover,  there exists a diagonal matrix $D$ and a partial permutation matrix $P$ such that  $\Psi = RM\cdot DP$.
\end{theorem}

\section{Experimental results}

To demonstrate that the ACS algorithm solves the ACS problem in practice, we train ACS networks on synthetic and natural image patches. We use $16 \times 16$ image patches which are compressed by an i.i.d. gaussian measurement matrix before ACS sees them. Unless otherwise stated we use a compression factor of 2; that is, the 256 dimensional patches were captured by 128 measurements sent to the ACS circuit (current experiments are successful with a compression factor of 10).  
The feature sign algorithm developed in \cite{lee2007} is used for inference of $\mathbf b$ in (\ref{ACSobjective}).  After the inference step, $\Theta$ is updated using gradient decent in (\ref{ACSobjective}).  The matrix $\Theta$ is initialized randomly and renormalized to have unit length columns after each learning step. Learning is performed until the ACS circuit converges on a sparsity basis for the compressed space.  

To assess whether the sparse representations formed by the ACS circuit are representations of the original data, we estimate a reconstruction matrix $RM$ as in (\ref{RMeqn}) by correlating a set of 10,000 uncompressed image patches with their encodings $\mathbf{b}$ in the ACS circuit.  Using $RM$ and the ACS circuit, we reconstruct original data from compressed data.  Reconstruction performance is evaluated on a test set of 1000 image patches by computing the signal-to-noise ratio of the reconstructed signals ${\mathbf{\widehat{x}}}$:   $SNR = 10 \log_{10}\left({ \langle || \mathbf{x} ||_2^2 \rangle_{X} \over  \langle || \mathbf{x} - \widehat{\mathbf{x}}||_2^2 \rangle_{X} }\right)$.
For comparison, we also performed CS using the feature sign algorithm to solve (\ref{Lasso}) using a fixed sparsity basis ${\Psi}$ and reconstruction given by $\mathbf{\widehat{x}} = \Psi \mathbf{\widehat{b}}$.

\begin{figure}
	\begin{minipage}[t]{0.33\linewidth}
	\centering
	\includegraphics[width=\linewidth]{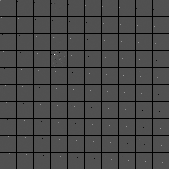}
	(a)
	\end{minipage}
	\begin{minipage}[t]{0.33\linewidth}
	\centering
	\includegraphics[width=\linewidth]{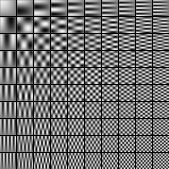}
	(b)
	\end{minipage}
	\begin{minipage}[t]{0.33\linewidth}
	\centering
	\includegraphics[width=\linewidth]{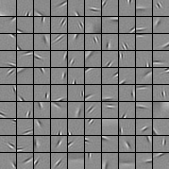}
	(c)
	\end{minipage}
	\label{synrm}
	\caption{Subsets of the reconstruction matrices $RM$ for the ACS networks trained on synthetic sparse data generated using bases (a) standard 2D,  (b) 2D DCT,  (c) learned by sparse coding on natural images. The components of RM in (a) and (b) are arranged by spatial location and spatial frequency respectively to help with visual interpretation.}
\end{figure}

\begin{figure}[t]
	\begin{minipage}[t]{0.5\linewidth} 
	\centering
	(a) \includegraphics[width=.8\linewidth]{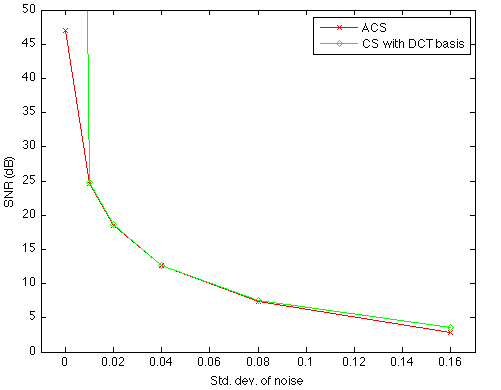}

	\vspace{.5cm}

	(c) \includegraphics[width=.8\linewidth]{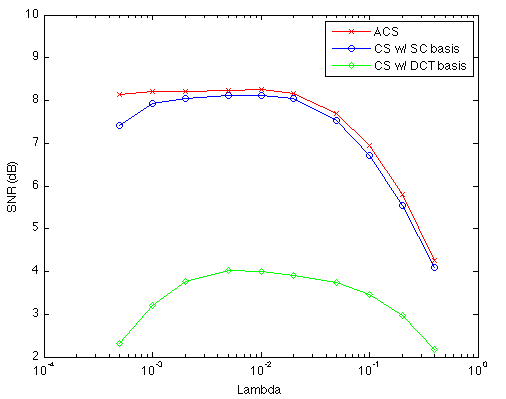}
	\end{minipage}
	\begin{minipage}[t]{0.5\linewidth}
	\centering
	(b) \includegraphics[width=.8\linewidth]{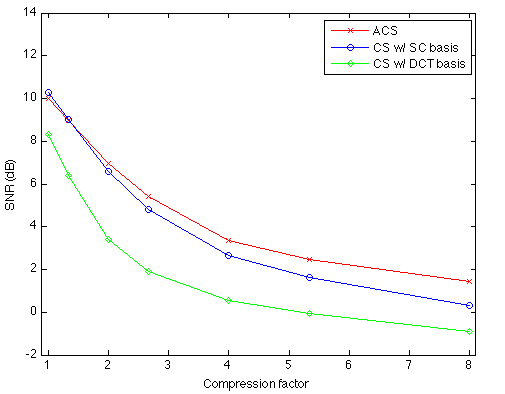}

	\vspace{.5cm}

	(d) \includegraphics[width=.8\linewidth]{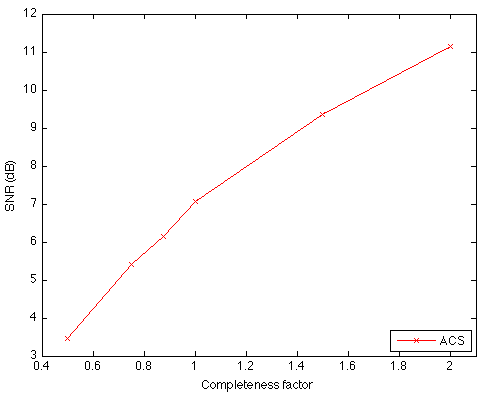}
	\end{minipage}
	\label{SNR}
	\caption{Mean SNR of reconstructions. (a) compares ACS performance to CS performance with true generating basis (DCT) for synthetic images with increasing amounts of gaussian noise. (b) and (c) compare the performances of ACS, CS with a basis learned by sparse coding on natural images and CS with the DCT basis. Performances plotted against the compression factor (b) and the value of $\lambda$ used for encoding.  (d) shows ACS performance on natural images vs. the completeness factor.}
\end{figure}

\textbf{Synthetic Data}:  To assess ACS performance on data of known sparsity we first generate synthetic image patches with sparse underlying structure in known bases.  We test with three different bases: the standard 2D basis (i.e. single pixel images), the 2D DCT basis, and a Gabor-like basis learned by sparse coding on natural images.  We generate random sparse binary vectors with $k = 8$, multiply these vectors by the chosen basis to get images, and then compress these images to half their original lengths to get training data.  For each type of synthetic data, a separate ACS network is trained with $\lambda = .1$ and reconstruction matrix $RM$ is computed. The $RM$ corresponding to each generating basis type is shown in Figure 2(a)-(c).  We can see that RM closely resembles a permutation of generating basis as predicted by Theorem~\ref{mainPsiDiagPermThm}.
The mean SNR of the reconstructed signals in each case is 34.05 dB, 47.05 dB, and 36.38 dB respectively.  Further, most ACS encodings are exact in the sense that they exactly recovered the components used to synthesize the original image.  Specifically, for the DCT basis 95.4\% of ACS codes have the same eight active basis vectors as were used to generate the original image patch. Thresholding to remove small coefficients (coring) makes it 100\%.  

To explore how ACS performs in cases where the signals cannot be modeled exactly with sparse representations, we generate sparse synthetic data ($k = 8$) with the 2D DCT basis and add gaussian noise.  Figure 3(a) compares reconstruction fidelity of ACS and CS for increasing levels of noise.  For pure sparse data (noise $\sigma^2 = 0$) CS outperforms ACS significantly.  Without noise, CS is limited by machine precision and reaches a mean SNR which is off the chart at 308.22 dB whereas ACS is limited by inaccuracies in the learning process as well as inaccuracies in computing $RM$.  For a large range of noise levels CS and ACS performance become nearly identical.  For very high levels of noise CS and ACS performances begin to diverge as the advantage of knowing the true sparsity basis becomes apparent again.

\begin{figure}
	\hspace{.03\linewidth} 
	\begin{minipage}[t]{0.4\linewidth} 
	\centering
	(a)
	\includegraphics[width = .8\linewidth]{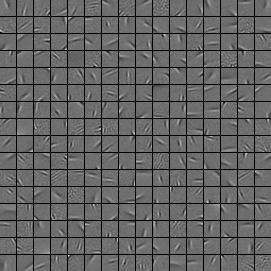}
	\end{minipage}
	\hspace{.1\linewidth} 
	\begin{minipage}[t]{0.4\linewidth}
	\centering
	(b)
	\includegraphics[width=.8\linewidth]{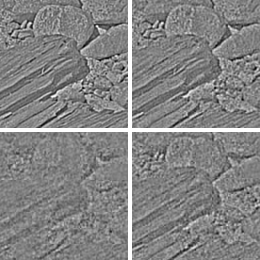}
	\end{minipage}
	\label{fig:ACS_RM}
	\caption{(a) $RM$ for an ACS network trained on natural images with compression factor of 2, (b) ACS reconstruction of a $128 \times 128$ image using increasing compression factors.  Clockwise from the top left: the original image, ACS with compression factors of 2, 4, and 8.}
\end{figure}

\textbf{Natural Images}:  Natural image patches have sparse underlying structure in the sense that they can be well approximated by sparse linear combinations of fixed bases, but they cannot be exactly reconstructed at a level of sparsity required by the theorems of CS and ACS. Thus, CS and ACS cannot be expected to produce exact reconstructions of natural image patches.  To explore the performance of ACS on natural images we train ACS models on compressed image patches from whitened natural images. 
The $RM$ matrix for an ACS network using the default compression factor of 2 is shown in Figure 4(a).

Next we explore how the fidelity of ACS reconstructions varies with the compression factor.  Figure 4(b) shows an entire image portion reconstructed patch-wise by ACS for increasing compression factors.  Figure 3(b) compares the SNR of these reconstructions to CS reconstructions.  Since there is no true sparsity basis for natural images, we perform CS either with a dictionary learned from uncompressed natural images using sparse coding or with the 2D DCT.  Both the ACS sparsity basis and sparse coding basis used with CS are learned with $\lambda$ fixed at .1 in eq. (\ref{SCObjFunc}).   3(b) demonstrates that CS performs much better with the learned dictionary than with the standard 2D DCT. Further, the plot shows that ACS reconstructions produces slightly higher fidelity reconstructions than CS.  However, the comparison between CS and ACS might be confounded by the sensitivity of these algorithms to the value of $\lambda$ used during encoding.

In the context of CS, there is a sweet spot for the sparsity of representations.  More sparse encodings have a better chance of being accurately recovered from the measurements because they obey conditions of the CS theorems better.  At the same time, these are less likely to be accurate encodings of the original signal since they are limited to fewer of the basis vectors for their reconstructions.  As a result, reconstruction fidelity as a function of $\lambda$ has a maximum at the sweet spot of sparsity for CS (decreasing the value of $\lambda$ leads to sparser representations).  Values of $\lambda$ below this point produce representations that are not sparse enough to be accurately recovered from the compressed measurements, while values of $\lambda$ above it produce representations that are too sparse to accurately model the original signal even if they could be accurately recovered.

To explore how the performance of CS and ACS depends on the sparseness of their representations, we vary the value of $\lambda$ used while encoding.  Figure 3(c) compares ACS, CS with a sparse coding basis, and CS with the 2D DCT basis.  Once again we see that ACS performs slightly better than CS with a learned dictionary, and much better than CS with the DCT basis.  However, the shape of the curves with respect to the choice of $\lambda$ while encoding suggests that our choice of value for $\lambda$ while learning (.1 for both ACS and the sparse coding basis used with CS) may be suboptimal.  Additionally, the optimal value of $\lambda$ for CS may differ from the optimal value of $\lambda$ for ACS.  For these reasons, it is unclear if ACS exceeds the SNR performance of CS with dictionary learning when in the optimal regime for both approaches.  Most likely, as 3(b) suggests, their performances are not significantly different.  However, one reason ACS might perform better is that learning a sparsity basis in compressed space tunes the sparsity basis with respect to the measurement matrix whereas performing dictionary learning for CS estimates the sparsity basis independently of the measurement matrix.  Additionally, having its sparsity basis in the compressed space means that ACS is more efficient in terms of runtime than dictionary learning for CS because the lengths of basis vectors are reduced by the compression factor.

\textbf{ACS in brain communication}:  When considering ACS as a model of communication in the brain, one important question is whether it works when the representational dimensions vary from region to region.  Typically in CS, the number of basis functions is chosen to equal the dimension of the original space. To demonstrate how ACS could model the communication between regions with different representation dimensions, we train ACS networks whose encoding dimensions are larger or smaller than the dimension of the original space (overcomplete or undercomplete).  As shown in figure 3(d), the reconstruction fidelity decreases in the undercomplete case because representations in that space either have fewer total active coding vectors or are significantly less sparse.  Interestingly, the reconstruction fidelity increases in the overcomplete case.  We suspect that 
this gain from overcompleteness also applies in standard CS with an overcomplete dictionary, but this has not been tested so far.

\begin{figure}[h]
	\centering
	\includegraphics[width = .9\linewidth]{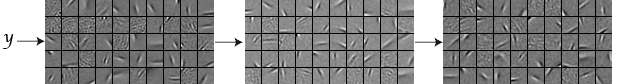}
	\caption{A subset of $RM$ from each stage of our multistage ACS model.}
	\label{fig:Multistage}
\end{figure}

Another issue to consider for ACS as a model of communication in the brain is whether signal fidelity is preserved through repeated communications.  To investigate this question we simulated multiple stages of communication using ACS.  In our model the input of compressed natural image patches is encoded as a sparse representation in the first region, transmitted as a compressed signal to a second region where it is encoded sparsely, and compressively transmitted once again to a third region that performs the final encoding.  Obviously, this is a vacuous model of neural computation since there is little use in simply retransmitting the same signal.  A meaningful model of cortical processing would involve additional local computations on the sparse representations before retransmission.  However, this basic model can help us explore the effects of repeated communication by ACS.  Using samples from the uncompressed space, we compute $RM$ for each stage just as for a single stage model.  Figure \ref{fig:Multistage} shows subsets of the components of $RM$ for each stage.  Notice that meaningful gabor-like structure is preserved between stages.  


\section{Discussion} 

In this paper, we propose ACS, a new algorithm for learning meaningful sparse representations of compressively sampled signals without access to the full signals. Two crucial differences set ACS apart from traditional CS. First, the ACS coding circuit is formed by unsupervised learning on subsampled signals and does not require knowledge of the sparsity basis of the signals nor of the measurement matrix used for subsampling. Second, the information in the fully trained ACS coding circuit is insufficient to reconstruct the original signals. To assess the usefulness of the representations formed by ACS, we developed a second estimation procedure that probes the trained ACS coding circuit with the full signals and correlates signal with encoding. Similarly to the electrophysiological approach of computing receptive fields, we computed a reconstruction matrix $RM$.
Theorem~4.2 proves that after convergence, ACS produces representations of the full data and that the estimation procedure finds a reconstruction matrix which can reproduce the full data.  Further, our simulation experiments revealed that the $RM$ matrix contained smooth receptive fields resembling oriented simple cells (Figures 2 and 4), suggesting that the ACS learning scheme can explain the formation of receptive fields even when the input to the cell population is undersampled (and thus conventional sparse coding would falter).  In addition, the combination of ACS circuit and $RM$ matrix can be used in practice for data compression and be directly compared with traditional CS.  Interestingly, ACS is fully on par with CS in terms of reconstruction quality (Figure 3). At the same time it is both flexible and stackable, and it works in overcomplete and undercomplete cases.

The recent work on BCS \cite{gleichmaneldar2010} addressed a similar problem where the sparsity basis of compressed samples is unknown. A main difference between BCS and ACS is that BCS aims for full reconstruction of the original signals from compressed signals whereas ACS does not.  As a consequence, BCS is generally ill-posed \cite{gleichmaneldar2010}, whereas ACS permits a solution, as we have shown. We have argued that full data reconstruction is not a prerequisite for communication between brain regions. However, note that ACS can be made a full reconstruction algorithm if there is limited access to uncompressed signal. Thus, neither ACS nor practical applications of BCS are fully blind learning algorithms, as both rely on further constraints \cite{gleichmaneldar2010} inferred from the original data. An alternative to ACS / BCS for introducing learning in CS was to adapt the measurement matrix to data \cite{elad2007,weissetal2007}. 

The engineering implications of ACS merit further exploration.  In particular, our compression results with overcomplete ACS indicate that the reconstruction quality was significantly higher than with standard CS.  Additionally, the unsupervised learning with ACS may have advantages in situations where access to uncompressed signals is limited or very expensive to acquire.  With ACS it is possible to do the heavy work of learning a good sparsity basis entirely in the compressed space and only a small number of samples from the uncompressed space are required to reconstruct with $RM$.

Perhaps the most intriguing implications of our work concern neurobiology.  Our results clearly demonstrate that meaningful sparse representations can be learned on the far end of wiring bottlenecks, fully unsupervised, and without any knowledge of the subsampling scheme. In addition, ACS with overcomplete or undercomplete codes suggests how sparse representations can be communicated between neural populations of different sizes.  From our study, we predict that firing patterns of neurons sending long-range axons might be less sparse than those involved in local connectivity, a hypothesis that could be experimentally verified.  It is intriguing to think that the elegance and simplicity of compressive sampling and sparse coding could be exploited by the brain.

\bibliographystyle{plain}
\bibliography{acs}

\end{document}